\documentclass[12pt,preprint]{aastex6}
\pdfoutput=1
\usepackage{amssymb}
\usepackage{amsmath}
\usepackage{color}
\usepackage{tabularx}
\usepackage{longtable}
\usepackage{txfonts}
\usepackage{wasysym}
\usepackage{url}
\usepackage{geometry}

\begin{document}
\newcommand{\x}{$\times$}
\newcommand{\lr}{$\longrightarrow$}
\newcommand{\ic}{^c}

\title{A new astrobiological model of the atmosphere of Titan}

\author{K. Willacy}
\affil{MS 169-507, Caltech/Jet Propulsion Laboratory, 4800 Oak  Grove Drive,  Pasadena, CA 91109}
\email{Karen.Willacy@jpl.nasa.gov}
\and
\author{M. Allen\altaffilmark{1}}
\affil{Caltech/Jet Propulsion Laboratory, 4800 Oak Grove Drive,  Pasadena, CA 91109}
\and
\author{Y. Yung}
\affil{Division of Geological and Planetary Science, California Institute of Technology, Pasadena, CA 91125}
\altaffiltext{1}{Division of Geological and Planetary Science, California Institute of Technology, Pasadena, CA 91125}

\begin{abstract}
We present results of an investigation into the formation of
nitrogen-bearing molecules  in the atmosphere of Titan.  We extend a
previous model \citep{li14,li15} to cover the region below the
tropopause, so the new model treats the atmosphere from Titan's
surface to an altitude of 1500 km.  We
consider the effects of condensation and sublimation using a
continuous, numerically stable method.  This is coupled with
parameterized treatments of the sedimentation of the aerosols and their condensates,
and the formation of haze particles.  These processes affect the abundances of heavier species such as
the nitrogen-bearing molecules, but have less effect on the abundances of
lighter molecules.  Removal of
molecules to form aerosols also plays a role in
determining the mixing ratios, in particular of HNC, HC$_3$N and HCN.
We find good agreement with the recently detected mixing ratios of
C$_2$H$_5$CN, with condensation playing an important role in
determining the abundance of this molecule below 500 km.   Of particular interest is
the chemistry of acrylonitrile (C$_2$H$_3$CN) which has been suggested
by \cite{slc15} as a molecule that could form biological
membranes in an oxygen-deficient environment.  With  the inclusion of 
haze formation we find good agreement of our model predictions of acrylonitrile with
the available observations.

\end{abstract}


\section{Introduction}
A major goal of planetary exploration is to obtain a fundamental understanding
of planetary environments, both as they are currently and as they were in the past.  
This knowledge
can be used to explore the questions of (a) how conditions for planetary habitability
arose and (b) the origins of life.  Titan is a unique object of study in this quest.  Other
than Earth itself, and Pluto \citep[which has also been observed to have
photochemically produced haze;][]{stern15,gladstone16}, 
Titan is the only solar system body demonstrated to have
complex organic chemistry occurring today.  Its atmospheric properties---
(1) a thick N$_2$ atmosphere, (2) a reducing
atmospheric composition, (3) energy sources for driving disequilibrium
chemistry and (4) an aerosol layer for shielding the surface from
solar UV radiation---suggest it is a counterpart of the early Earth, before the latter's
reducing atmosphere was eradicated by the emergence and evolution of life
\citep{ct99,lunine05,lm08}.

A significant number of photochemical models have been developed to
investigate the distribution of hydrocarbons in Titan's atmosphere
\citep{strobel74,yap84,lara96,wa04,delahaye08,lavvas08a,lavvas08b,krasnopolsky09}.  
Recently, more constraints have been placed on the abundance of
hydrocarbons and nitriles in the mesosphere of Titan (500 -- 1000 km)
from Cassini/UVIS stellar occultations \citep{koskinen11,kammer13}.
 In combination with the updated version of Cassini/ CIRS
limb view \citep{vinatier10}, the complete profiles of C$_2$H$_2$, C$_2$H$_4$,
C$_6$H$_6$, HCN, HC$_3$N are revealed for the first time. C$_3$ compounds,
including C$_3$H$_6$, were modeled by \cite{li15}, and the agreement
with observations \citep{nixon13} is satisfactory.  
The chemistry of many of these nitrogen molecules has recently been 
modeled by \cite{loison15}.

In this paper we introduce our updated Titan chemical model that
includes the formation of such potentially astrobiologically important
molecules as acrylonitrile.  In addition to the usual gas phase
chemistry, it also includes a numerically stable treatment of the
condensation and sublimation, allowing the formation and destruction
of ices in the lower atmosphere to be tracked.  Haze formation is also
included in a parameterized fashion, allowing for the permanent
removal of molecules from the atmosphere.  We present here
the effects of condensation on the nitrogen chemistry.  The
interaction of hydrocarbons and nitrile species in the condensed
phase is complex and is beyond the scope of this paper \citep[see, for
example, Figures 1 and 2 of][]{anderson16}.

We begin with describing our updated model and in particular our
treatment of condensation and sublimation (Section~\ref{sec:model}).
We use this updated model to consider the chemistry in Titan's
atmosphere from the surface of the moon to an altitude of 1500 km.  We
explore how condensation processes and haze formation affect the
predicted gas phase abundances of observable molecules
(Section~\ref{sec:results}).   We also consider where the
condensates form within the atmosphere (Section~\ref{sec:cond}).
Section~\ref{sec:conc} presents our conclusions.

\section{The Model}\label{sec:model}

We use the Caltech/JPL photochemical model \citep[KINETICS;][]{allen81}
with a recently updated
chemical network, and with the addition of condensation and
sublimation  processes to explore the atmospheric chemistry of Titan.  
The 1-D model solves the mass continuity equation from the surface of
Titan to 1500 km altitude:
\begin{equation}
\frac{\partial n_i}{\partial t} + \frac{\partial \psi_i}{\partial z} = P_i - L_i
 \end{equation}
where $n_i$ is the number density of species $i$, and $P_i$ and $L_i$ 
are its chemical production and loss rates respectively.  $\psi_i$ is
the vertical flux of $i$ calculated from 
\begin{equation}
\psi_i = - \frac{\partial n_i}{\partial z}(D_i + K_{zz}) - n_i \left( \frac{D_i}{H_i} + \frac{K_{zz}}{H_a} \right) - n_i \frac{\partial T}{\partial z} \frac{(1+\alpha_i)D_i + K_{zz}}{T}
\end{equation}
where $D_i$ and $H_i$ are the molecular diffusion coefficient and the 
scale height for species $i$ respectively, $H_a$ is the atmospheric
scale height, $\alpha_i$ is the thermal diffusion coefficient of
species $i$, $T$ is the temperature and $K_{zz}$ is the eddy diffusion 
coefficient.  The eddy diffusion coefficient used here is taken from 
\cite{li15} and can be summarized as
\begin{equation}
log K_{zz}(z) = \left\{
\begin{array}{ll}
log(3 \times 10^3) , &  z < z_1\\
\\
log(3 \times 10^3) \dfrac{z_2 - z}{z_2 -z_1} + log(2 \times 10^7) \dfrac{z - z_1}{z_2 - z_1} , &  z_1 \leq z < z_2 \\
\\
log(2 \times 10^7) \dfrac{z_3 - z}{z_3 - z_1} + log(2 \times 10^6) \dfrac{z - z_2}{z_3 - z_2} , &  z_2 \leq z < z_3\\
\\
log(2 \times 10^6) \dfrac{z_4 - z}{z_4 - z_3} + log(4 \times 10^8) \dfrac{z - z_3}{z_4 - z_3} , &  z_3 \leq z < z_4\\
\\
log(4 \times 10^8) , &  z \geq z_4
\end{array}
\right.
\end{equation}

\noindent The atmospheric density and temperature profiles are also taken 
from \cite{li15}, and are based on the T40 Cassini flyby
\citep{westlake11}. 

\begin{figure}
\includegraphics[width=0.5\linewidth]{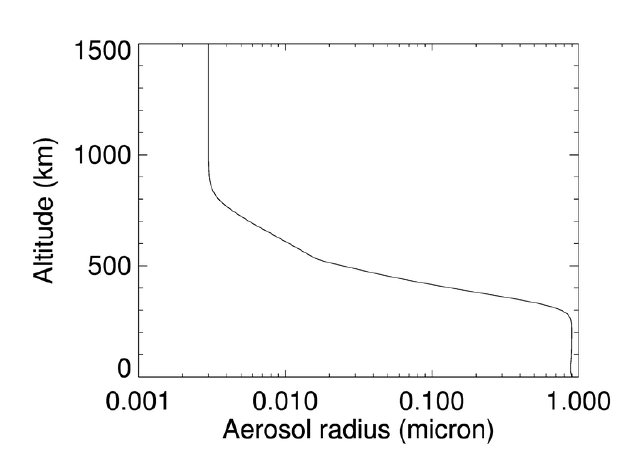}
\includegraphics[width=0.5\linewidth]{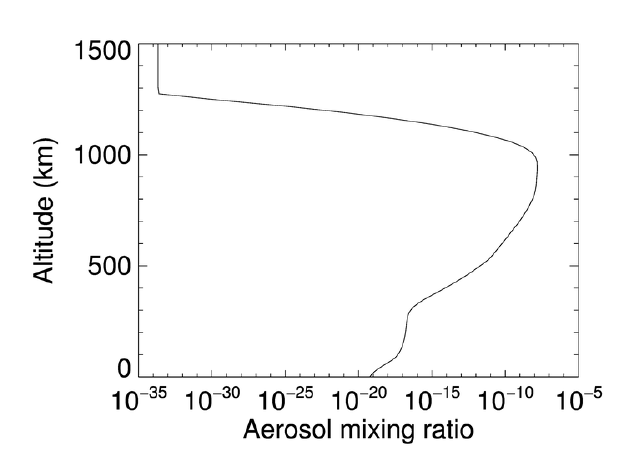}
\caption{\label{fig:aerosol}Aerosol properties from
  \cite{lavvas10} derived from a microphysical model validated against
  DISR observations. {\it left:} The mean radius of
  particles.  {\it right:} The mixing ratio of particles.}
\end{figure}

Aerosols are included in our model, both for the absorption of UV
radiation and to provide surfaces onto which molecules can condense. 
The aerosol properties are from \cite{lavvas10} who derived them
from a microphysical model 
validated against Cassini/Descent Imager Spectral Radiometer (DISR) observations.  
Their results provide the mixing
ratio and surface area of aerosol particles as a function of
altitude (Figure~\ref{fig:aerosol}).  To calculate the absorption of UV 
by dust we assume absorbing aerosols with
extinction cross-sections that are independent of wavelength
\citep{li14,li15}. 

\subsection{Boundary Conditions}

The lower boundary of our model is the surface of Titan and the upper
boundary is at 1500 km.   For H and H$_2$ the flux at the lower
boundary is zero and at the top of the atmosphere these molecules are
allowed to escape with velocities of 2.4
$\times$ 10$^4$  cms$^{-1}$ and 6.1 $\times$ 10$^{3}$
cms$^{-1}$ respectively (equivalent to fluxes of 3.78 $\times$
  10$^8$ H atoms cm$^{-2}$ and 6.2 $\times$ 10$^9$
H$_2$ molecules cm$^{-2}$). For all other gaseous species the
concentration gradient at the lower boundary is assumed to be zero,
while they have zero flux at the top boundary.  
Observations suggest that CH$_4$ can escape from the top of the
atmosphere by sputtering \citep{delahaye07} but the same effect can be
generated in models by applying a larger eddy diffusivity
\citep{li15,li14,yelle08} which is the approach we have taken here.
Condensed species have zero flux at both the upper and lower boundaries. 

Table~\ref{tab:molecules} provides a list of the molecules in 
our model. The mixing ratio of N$_2$ is set according to the
  observational data and held fixed, with values below 50 km taken from
  the Huygens observations \citep{niemann05} and above 1000 km from
  Cassini/UVIS data \citep{kammer13}. Between 50 and 1000 km the
  mixing ratio is assumed to be 0.98.
The mixing ratio of CH$_4$ is fixed to the observed (super-saturated) values 
\citep{niemann10} below the tropopause and allowed to vary above this.

\begin{table}
\begin{tabularx}{\linewidth}{l X}
\hline
Family & Molecule \\
\hline
& H, H$_2$\\
& \\
hydrocarbons & C  CH  CH$_2$ $^3$CH$_2$  CH$_3$  CH$_4$  C2  C$_2$H  C$_2$H$_2$ 
               C$_2$H$_3$  C$_2$H$_4$  C$_2$H$_5$  C$_2$H$_6$  C$_3$
               C$_3$H  C$_3$H$_2$  C$_3$H$_3$  
               C$_2$CCH$_2$  CH$_3$C$_2$H  C$_3$H$_5$  C$_3$H$_6$
               C$_3$H$_7$  C$_3$H$_8$  C$_4$H  C$_4$H$_2$  
               C$_4$H$_3$  C$_4$H$_4$  C$_4$H$_5$  1-C$_4$H$_6$
               1,2-C$_4$H$_6$  1,3-C$_4$H$_6$  C$_4$H$_8$
               C$_4$H$_9$  
               C$_4$H$_{10}$  C$_5$H$_3$  C$_5$H$_4$  C$_6$H
               C$_6$H$_2$  C$_6$H$_3$  C$_6$H$_4$  C$_6$H$_5$
               $l$-C$_6$H$_6$  
               C$_6$H$_6$  C$_8$H$_2$\\
& \\
nitrogen-molecules & N  NH  NH$_2$  NH$_3$  N$_2$H  N$_2$H$_2$
N$_2$H$_3$  N$_2$H$_4$ 
                     CN  HCN  HNC  H$_2$CN  CHCN  CH$_2$CN  CH$_3$CN
                     C$_2$H$_3$CN  
                     C$_2$H$_5$CN  C$_3$H$_5$CN  C$_2$N$_2$
                     HC$_2$N$_2$  C$_3$N  HC$_3$N  HC$_4$N  
                     CH$_3$C$_2$CN  H$_2$C$_3$N  C$_4$N$_2$  HC$_5$N
                     C$_6$N$_2$  CH$_2$NH  CH$_2$NH$_2$  
                     CH$_3$NH  CH$_3$NH$_2$ \\
& \\
condensed molecules & C$_2$H$_2\ic$ C$_2$H$_4\ic$  C$_2$H$_6\ic$  CH$_2$CCH$_2\ic$
CH$_3$C$_2$H$\ic$   
                      C$_3$H$_6\ic$  C$_3$H$_8\ic$  C$_4$H$_2\ic$  C$_4$H$_4\ic$
                      1-C$_4$H$_6$$\ic$ 1,2-C$_4$H$_6$$\ic$ 1,3-C$_4$H$_6$$\ic$
                      C$_4$H$_8\ic$  C$_4$H$_{10}\ic$  C$_5$H$_4\ic$  
                      $l$-C$_6$H$_6\ic$  C$_6$H$_6\ic$  HCN$\ic$  HNC$\ic$ CH$_3$CN$\ic$
                      C$_2$H$_3$CN$\ic$  C$_2$H$_5$CN$\ic$  
                      C$_3$H$_5$CN$\ic$  C$_2$N$_2\ic$  C$_4$N$_2\ic$  C$_6$N$_2\ic$
                      HC$_3$N$\ic$  HC$_5$N$\ic$  CH$_3$C$_2$CN$\ic$   
                      CH$_2$NH$\ic$  CH$_3$NH$_2\ic$  NH$_3\ic$
                      N$_2$H$_2\ic$ N$_2$H$_4\ic$  \\
\end{tabularx}
%\hline
\caption{\label{tab:molecules}The species included in the model.  A
  superscript of $\ic$ indicates the molecule is condensed.  }
\end{table}

\subsection{Condensation and Sublimation\label{sec:cond}}

Condensation occurs when the saturation ratio, $S$, of a molecule is greater than 1. 
S is defined as n($x$)/n$_{sat}$(s), where n($x$) is the
gas phase mixing ratio of species $x$ and n$_{sat}$($x$) is its saturated density
derived from the saturated vapor pressure.  For $S$ $<$ 1, condensation is
switched off and sublimation of any adsorbed molecules can occur.  
The abrupt change in behavior at $S$ = 1 can lead to numerical 
instabilities where the system oscillates between the condensation and
sublimation regimes. In previous Titan 
models various methods have been used to smooth out the transition 
and prevent such instabilities.  
 For example, \cite{yap84}
parameterized the condensation rate in terms of $S$:
\begin{equation}
\hbox{Loss rate} \propto - \frac{S-1}{S}
\end{equation}
This results in a relatively constant loss rate as a function of $S$.  
A more complicated expression was used by \cite{lavvas08a} 
to ensure that the loss rate increases with increasing saturation ratios:
\begin{equation}
\hbox{Loss rate} \propto -(S-1) \frac{exp(-0.5/ln(S+1)^2)}{ln(S+1)^2} \hbox{~~~for S $>$ 1}
\end{equation}
Other expressions that have been invoked include
%\begin{equation}
%\hbox{Loss rate} \propto -(1 0 1/S) (THIS IS WRONG)
%\end{equation}
%\citep{wa04} and
\begin{equation}
\hbox{Loss rate} \propto -ln S
\end{equation}
\citep{krasnopolsky09}.

Here we use a numerically stable method to determine the net 
condensation rate.  The rate at which molecules condense on to
a pre-existing aerosol particle is given by 
the collision rate with the particle:
\begin{equation}
k_{c} = \alpha_x \sigma v_x n(x) \hbox{~~~molecules s$^{-1}$} \label{eq:freeze}
\end{equation}
where $\alpha$ is the sticking coefficient of molecule $x$ (where $\alpha_x$ $\leq$ 1), 
$\sigma$ is the collisional cross-section of the particles, 
$v_x$ is the gas phase
velocity of $x$, and $n(x)$ is its number density.
For a pure ice the saturated vapor pressure is measured when the 
condensation and sublimation processes are in equilibrium.  In this scenario
\begin{equation}
k_c n_{sat}(x) = k_s \Theta_x \label{eq:sub}
\end{equation}
where $k_s$ is the sublimation rate and $\Theta_x$ is the surface 
coverage of molecule $x$.  In the case of a pure ice, $\Theta_x$ = 1, 
and hence the sublimation rate, $k_s$ = $k_c n_{sat}(x)$.  The net
condensation rate, $J_c$ is therefore
\begin{equation}
J_c = \alpha_x \sigma v_x [n(x) - n_{sat}(x)\Theta_x] \hbox{~~~molecules s$^{-1}$}
\end{equation}
 
When sublimation is taking place from a mixture of ices (rather than from pure ice) 
$\Theta$ will be less than 1 and the resulting gas phase abundance will
be lower than the saturated value.
$\Theta$ is calculated from
\begin{equation}
\Theta = n(x^c)/\Sigma_y n(y^c)
\end{equation}
where $n(x^c)$ is the number density of $x$ in the condensed phase,
$\Sigma_y n(y^c)$ is the total number density of all molecules condensed on to the
grain surface.  We assume that the ices are well-mixed, so that the
composition of the surface from which sublimation occurs reflects that
of the bulk of the ice.

\begin{figure}
\includegraphics[width=0.5\linewidth]{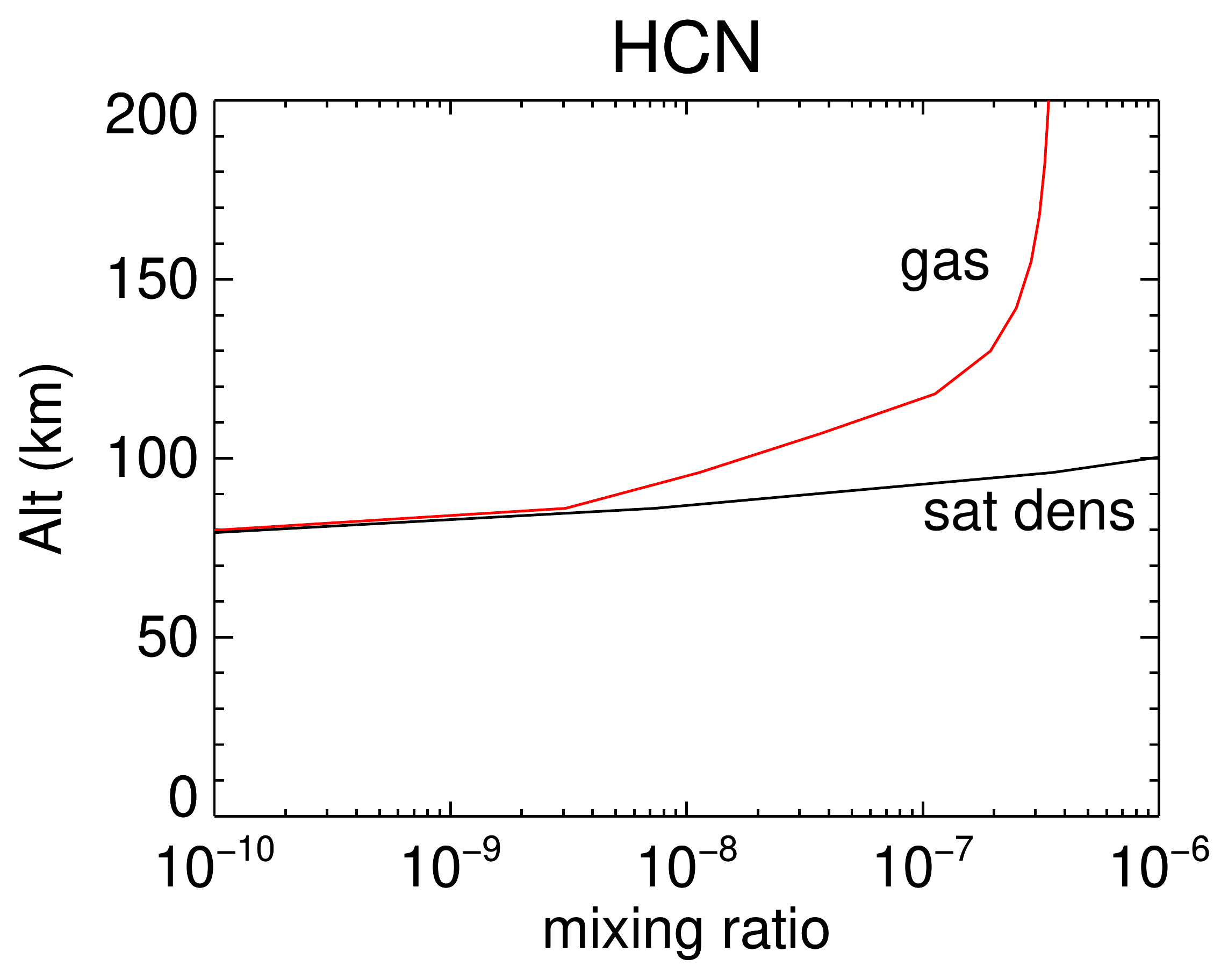}
\includegraphics[width=0.5\linewidth]{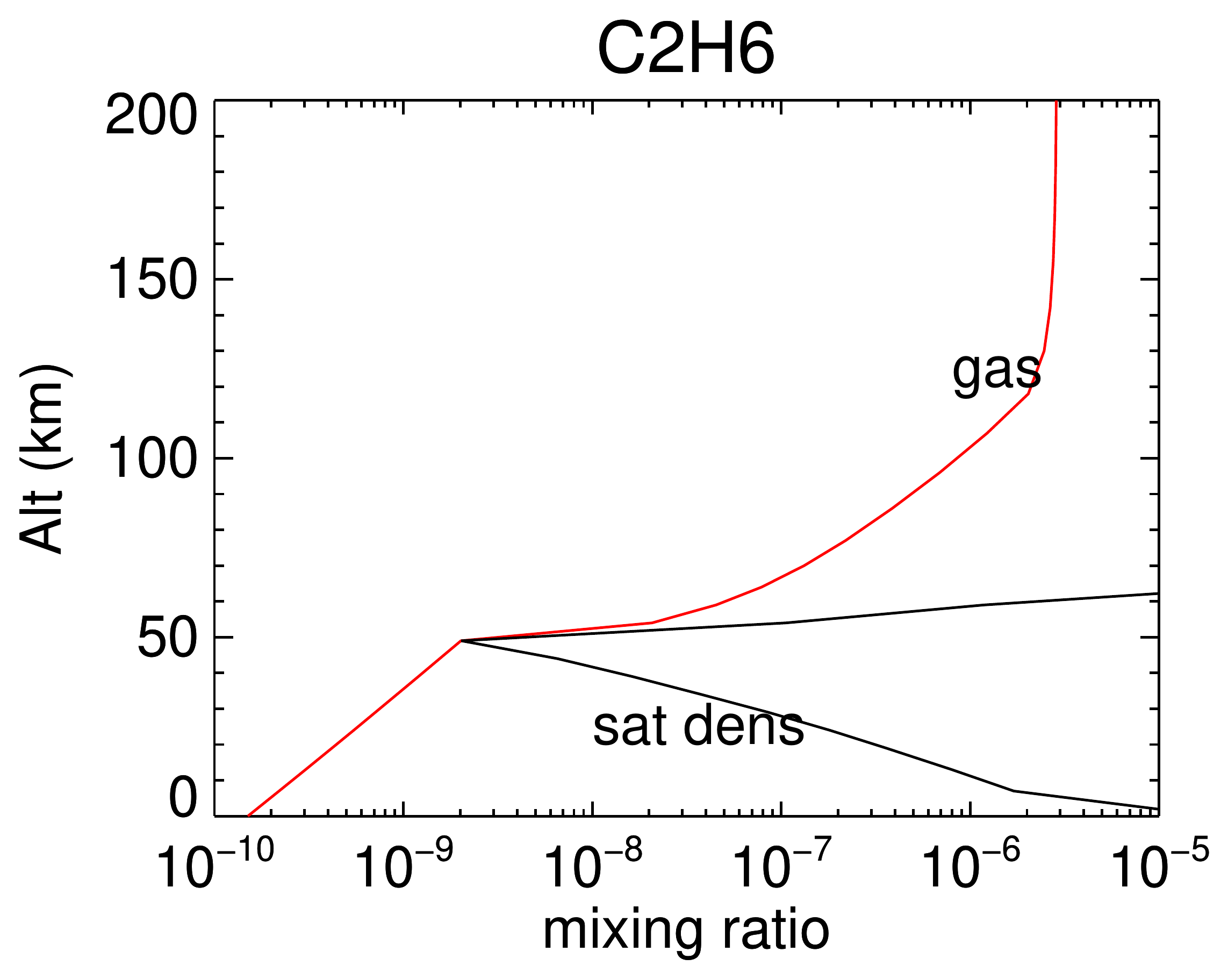}
\caption{\label{fig:comp}A comparison of the model results and the
  calculated mixing ratio under saturation conditions for (a) HCN and
  (b) C$_2$H$_2$. The saturated
  value (black line) is calculated assuming a pure ice and using the vapor pressure expressions
  given in Table~\ref{tab:ant}.  The model is shown in
  red.  The two lines coincide in the region where the model
  calculations reach the saturated value and deviate where the
  calculated mixing ratios are below the saturated values.}
\end{figure}

To determine the saturated densities used in this paper
we use the expressions for the saturated vapor pressure given in
Table~\ref{tab:ant}. 
The values from these fits are extrapolated as necessary to provide
saturation vapor pressures over a wider range of temperatures.   
Figure~\ref{fig:comp} compares the predicted mixing ratios of HCN and C$_2$H$_2$
with the value predicted directly from the saturated vapor pressure.  It can be seen that the model 
produces good agreement with the saturation vapor pressure in regions where the gas
is saturated.

\begin{table}
\centering
\scriptsize
\begin{tabular}{lccl}
\hline
Molecule & Expression for log P$_{sat}$& Temp range  & Notes \\
              &  (mmHg)                & (K) \\
\hline
CH$_4$ & 6.84570 -  435.6214/(T-1.639) & 91 -- 189 & \cite{yaws}\\ 
C$_2$H$_2$ & 6.09748 - (1644.1/T) + 7.42346 log(1000./T) & 80 -- 145 &\cite{moses92}\\
                   & 7.3147 - 790.20947/(T-10.141) & 192 -- 208 & \cite{lara96}\\
C$_2$H$_4$ & 1.5477 - 1038.1 (1/T - 0.011) + 16537./(1/T - 0.011)$^2$ & 77 -- 89 & \cite{moses92} \\
                  & 8.724 - 901.6/(T-2.555) & 89 --104 & \cite{moses92}\\
                  & 50.79 - 1703./T - 17.141 log(T) & 104 -- 120 & \cite{moses92}\\
                   & 6.74756 - 585./(T-18.18) & 120 -- 155 & \cite{moses92}\\
C$_2$H$_6$ & 10.01 - 1085./(T - 0.561) & 30 -- 90 & \cite{lara96} \\
                  & 6.9534 - 699.10608/(T-12.736) & 91 -- 305 & \cite{yaws}\\
CH$_3$C$_2$H & 6.78485 - 803.72998/(T-43.92) & 183 -- 267 &  \cite{yaws}\\
CH$_2$CCH$_2$ & 6.62555 - 684.69623/(T-55.658) & 144 -- 294  &  \cite{yaws}\\
C$_3$H$_6$ & 6.8196 - 785./(T-26.) & 161 -- 241 &  \cite{yaws}\\
C$_3$H$_8$ & 7.0189 - 889.8642/(T-15.916) & 85 -- 176 & \cite{yaws}\\
C$_4$H$_2$ & 5.3817 - 3300.5/T + 16.63415 log$_{10}$(1000./T) & 127--237 & \cite{lara96}\\
                  & 6.5326 - 761.68429/(T-74.732) & 237 -- 478  & \cite{yaws}\\
C$_4$H$_4$ & 6.6633 -  826.0438/(T-59.712) & 181 -- 454  & \cite{yaws}\\
1-C$_4$H$_6$ & 6.98198 - 988.75(T-39.99) & 205 -- 300  & \cite{yaws}\\
1,2-C$_4$H$_6$ & 6.99383 - 1041.117/(T-30.726) & 247 -- 303 & \cite{yaws}\\
1,3-C$_4$H$_6$ & 6.84999 - 930.546/(T-34.146) & 215 -- 287 & \cite{yaws}\\
C$_4$H$_8$ & 6.8429 -  926.0998/(T-33.) & 192 --  286 &  \cite{yaws}\\
C$_4$H$_{10}$ & 7.0096 - 1022.47681/(T-24.755) & 135 --  425  & \cite{yaws}\\
C$_5$H$_4$ & 7.986 - 1509.98716/(T-32.226) & 234 -- 367  & \cite{yaws}\\
l-C$_6$H$_6$ & 7.95508 - 1773.77625/(T-52.937) & 341 -- 449 & \cite{yaws}\\
C$_6$H$_6$ & 6.814 - 1090.43115/(T-75.852) & 233 -- 562 &  \cite{yaws}\\
NH$_3$ & 7.5874 - 1013.78149/(T-24.17) & 196 -- 405 &  \cite{yaws}\\
HCN & 11.41 - 2318./T & 132 -- 168 & \cite{lara96} \\
      & 8.0258 - 1608.28491/(T-286.893) & 260 -- 456 &  \cite{yaws}\\
HNC & 11.41 - 2318./T & 132 -- 168 & same as HCN \\
      & 8.0258 - 1608.28491/(T-286.893) & 260 -- 456 &  same as HCN\\
C$_2$N$_2$ & 6.9442 -  779.237/(T-60.078) & 146 -- 400 & \cite{yaws}\\
C$_4$N$_2$ & 8.269 - 2155./T  & 147 -- 384 & \cite{yaws}\\
C$_6$N$_2$ & 8.269 - 2155./T  & 147 -- 384 & same as C$_4$N$_2$ \\
HC$_3$N & 6.2249 -  714.01178/(T-101.55) & 214 --  315 & \cite{yaws}\\
HC$_5$N & 6.2249 -  714.01178/(T-101.55) & 214 -- 315 & same as HC$_3$N\\
C$_2$H$_3$CN & 7.8376 - 1482.7653/T-25.) & 189 -- 535 & \cite{yaws}\\
C$_2$H$_5$CN & 7.0414 - 1270.41907/(T-65.33) & 204 -- 564 & \cite{yaws}\\
C$_3$H$_5$CN & 7.0406 - 1617.87915/(T-34.032) & 186 -- 583 & \cite{yaws}\\
N$_2$H$_2$ & 7.8288 - 1698.58081/(T-43.21) & 270 -- 653 & same as N$_2$H$_4$\\
N$_2$H$_4$ & 7.8288 - 1698.58081/(T-43.21) & 270 -- 653 & \cite{yaws}\\
CH$_3$NH$_2$ & 7.3638 - 1025.39819/(T-37.938) & 180 -- 430 & \cite{yaws}\\
CH$_3$CN & 6.8376 -  995.2049/(T-80.494) & 266 -- 518 & \cite{yaws}\\
CH$_3$C$_2$CN & 6.2249 - 714.01178/(T-101.855) & 214 -- 315 &  \cite{yaws}\\
CH$_2$NH &  8.0913 - 1582.91077/(T-33.904) & 175 --512 & From \cite{yaws} value for CH$_3$OH\\
\hline
\end{tabular}
\caption{\label{tab:ant} Expressions used to calculate the saturated vapor pressures.  In the absence
of laboratory data we assume that the saturated vapor pressure of HNC is the same as HCN, and that
of N$_2$H$_2$ is the same as N$_2$H$_4$.  We follow \cite{loison15} in using the vapor pressure of
CH$_3$OH for CH$_2$NH and in using H$_3$CN for H$_5$CN.}
\end{table}

\subsection{\label{sec:sed}Sedimentation and Haze Formation}

We assume that the abundance, size and location of the
aerosol particles is fixed.  In reality the particles do not remain at the same altitude but
rather sediment out towards the surface of Titan, taking any condensates
with them.  To mimic this effect we have included a loss process for
condensed molecules which removes them from the model atmosphere with a rate 
coefficient of 10$^{-10}$ s$^{-1}$.  All condensed
species are assumed to be lost at the same rate.  The assumed size of this
reaction rate is somewhat arbitrary and to test the sensitivity of our results 
to its value we
also considered a loss rate of 10$^{-12}$
molecules s$^{-1}$.  Changing the rate was found to have no effect on the predicted
gas phase mixing ratios.

In addition to the condensation of ice or liquids onto existing aerosols,
molecules can also be incorporated into new or existing aerosols.  In
this scenario the molecules are then unavailable for return to the gas
via sublimation and are permanently removed from the gas
\citep{liang07}.   This process is simulated using rates that 
are proportional to the collision rates between aerosols (assuming mean radii
provided by \cite{lavvas10}) and molecules.
We simulate this by adding reactions that remove the molecules from the gas with 
\begin{equation}
\hbox{X} + haze = haze \hbox{~~~k = $\beta v \sigma n_g$ s$^{-1}$}
\label{eq:aero}
\end{equation}
where $n_g$ is the mixing ratio of aerosol particles, and 
$\beta$ is an efficiency factor ranging from 0.01 to 10 depending
on the molecule.  The value of $\beta$ was chosen for each molecule
to maximize the agreement of the models with the observations.  The
molecules to be removed in this way are HCN 
($\beta$ = 0.01), C$_2$H$_3$CN ($\beta$ = 0.1), HC$_3$N and HNC ($\beta$ = 10)
 C$_2$H$_5$CN ($\beta$ = 1). Other molecules
are assumed not to condense in this way -- for these molecules
agreement of the models with observations is sufficiently good 
without invoking an additional loss mechanism such as haze formation.

\section{Results}\label{sec:results}

\subsection{The Effect of Condensation Processes}

We present the results of three models with different assumptions
about the condensation and sublimation.  Model A is a gas phase only
model, with no condensation.  Model B includes condensation and
sublimation processes as outlined in Section~\ref{sec:cond}, and the
sedimentation of aerosol particles and their condensates.  Model C
extends Model B to include 
the removal of molecules from the gas by haze formation.  
The model parameters are summarized in
Table~\ref{tab:model}.

\begin{table}
\begin{tabular}{cccc}
\hline
Model & Condensation & Sedimentation & Haze \\
& & & formation \\
\hline
A & No & No & No \\
B & Yes & Yes & No \\
C & Yes & Yes & Yes \\
\hline
\end{tabular}
\caption{\label{tab:model}Summary of model assumptions.  Condensation
  and sublimation rates are discussed in
  Section~\ref{sec:cond}. Sedimentation and haze formation rates are
  discussed in Section~\ref{sec:sed}.}
\end{table}

%For most species the effect of condensation on the gas abundances can
%only be seen in the
%lower atmosphere (below 49 km).  Here the densities are high and the
%temperatures low enough for significant amounts of many species to be
%condensed out.   The gas phase abundance curves follow the vapor
%pressure curves in those regions of the atmosphere where the molecular
%abundance is sufficiently high (Figure~\ref{fig:comp}).

The largest effects are seen for the biggest molecules and in particular for those that
contain nitrogen.  The addition of sedimentation increases the
rate of removal of these species from the gas in the lower atmosphere and
improves agreement with the observations.   
However, some molecules are still found to be over-abundant. 
Further improvement is
achieved between 200 and 600 km for HCN, HNC, 
HC$_3$N and C$_2$H$_5$CN if these molecules are assumed to be incorporated
into haze particles. 

Below we discuss the chemistry of several species in more detail.

\subsection{Distribution of Nitrogen Molecules}

\begin{figure}
\includegraphics[width=0.5\linewidth]{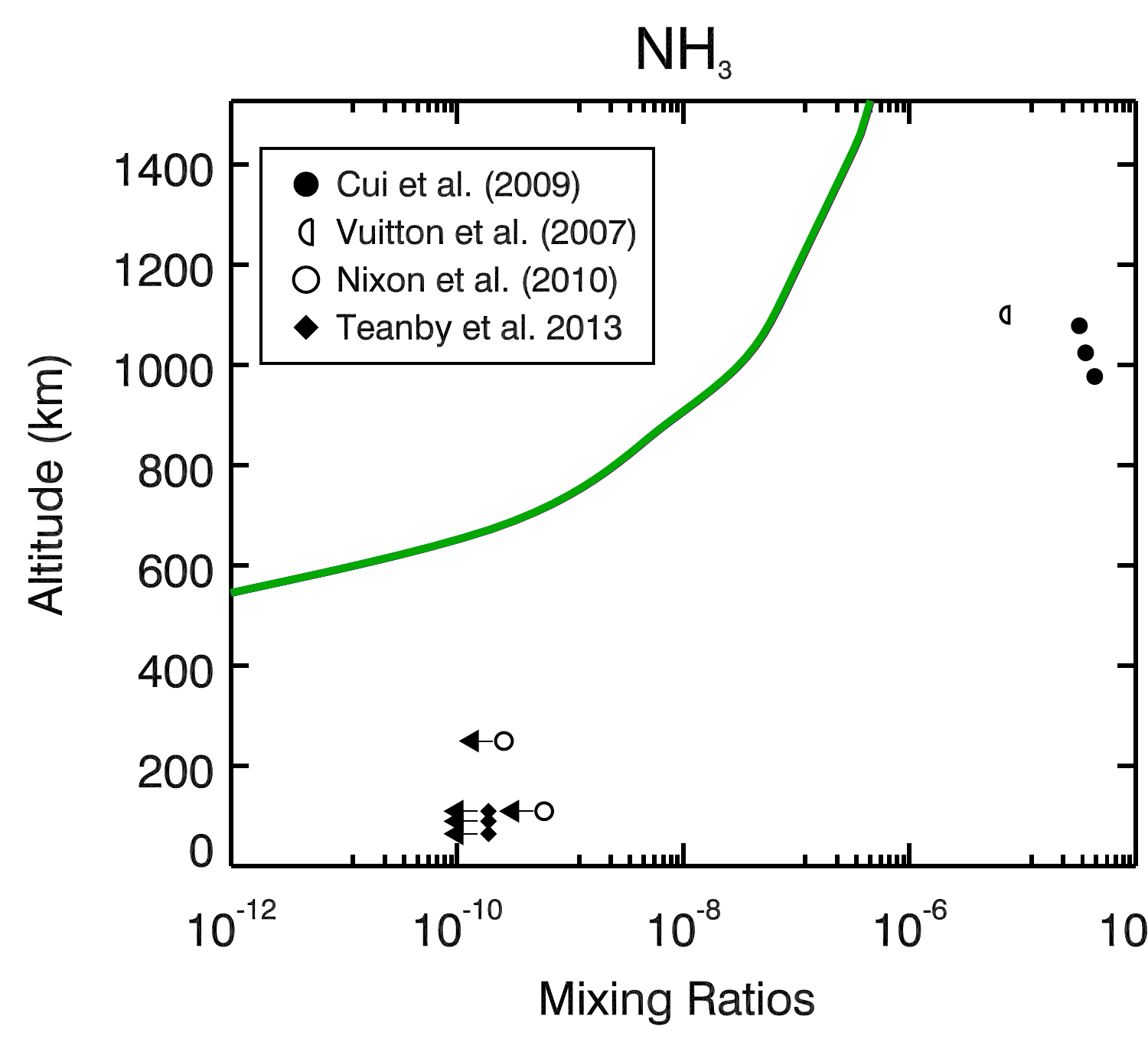}
\includegraphics[width=0.5\linewidth]{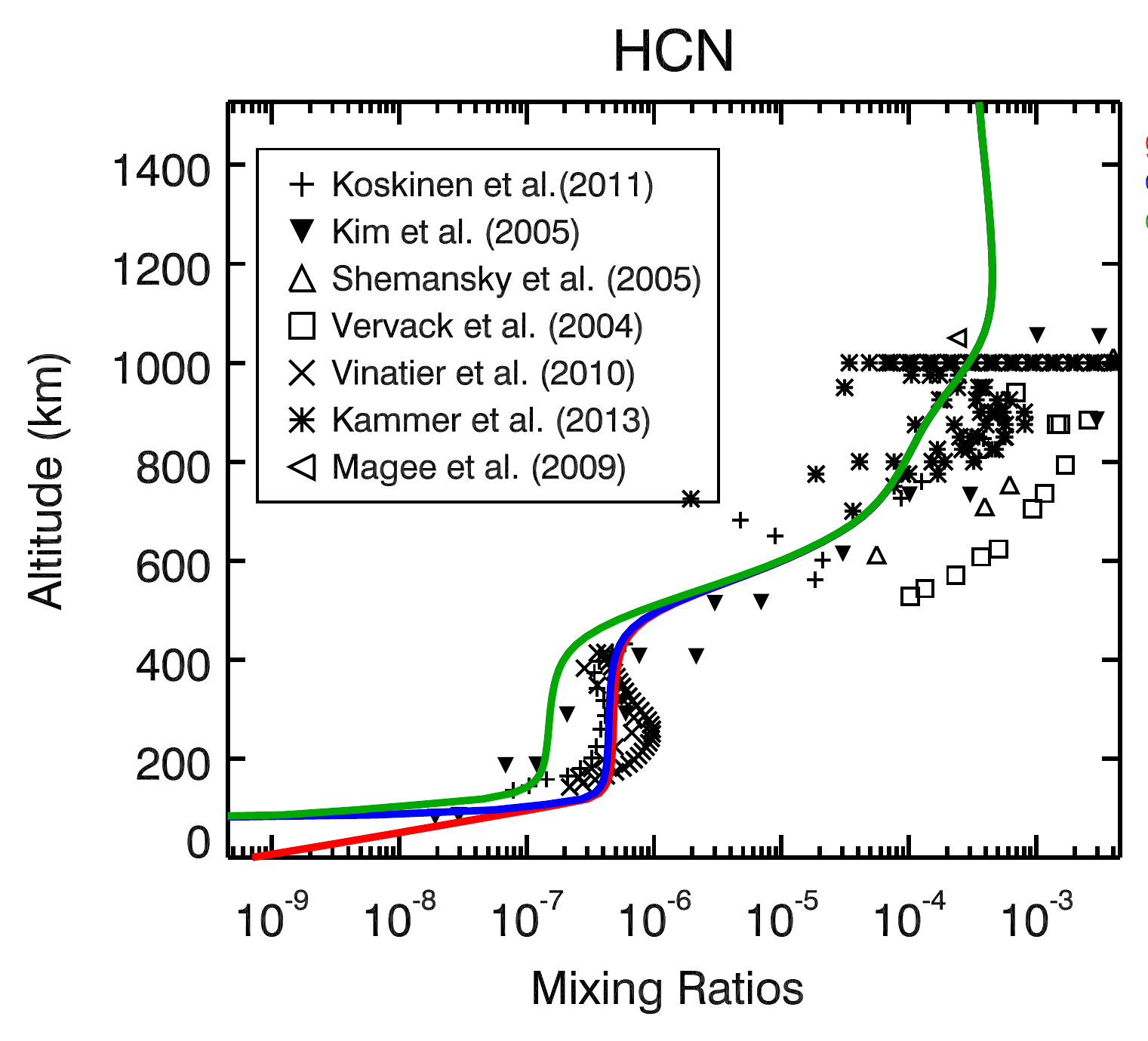}
\includegraphics[width=0.5\linewidth]{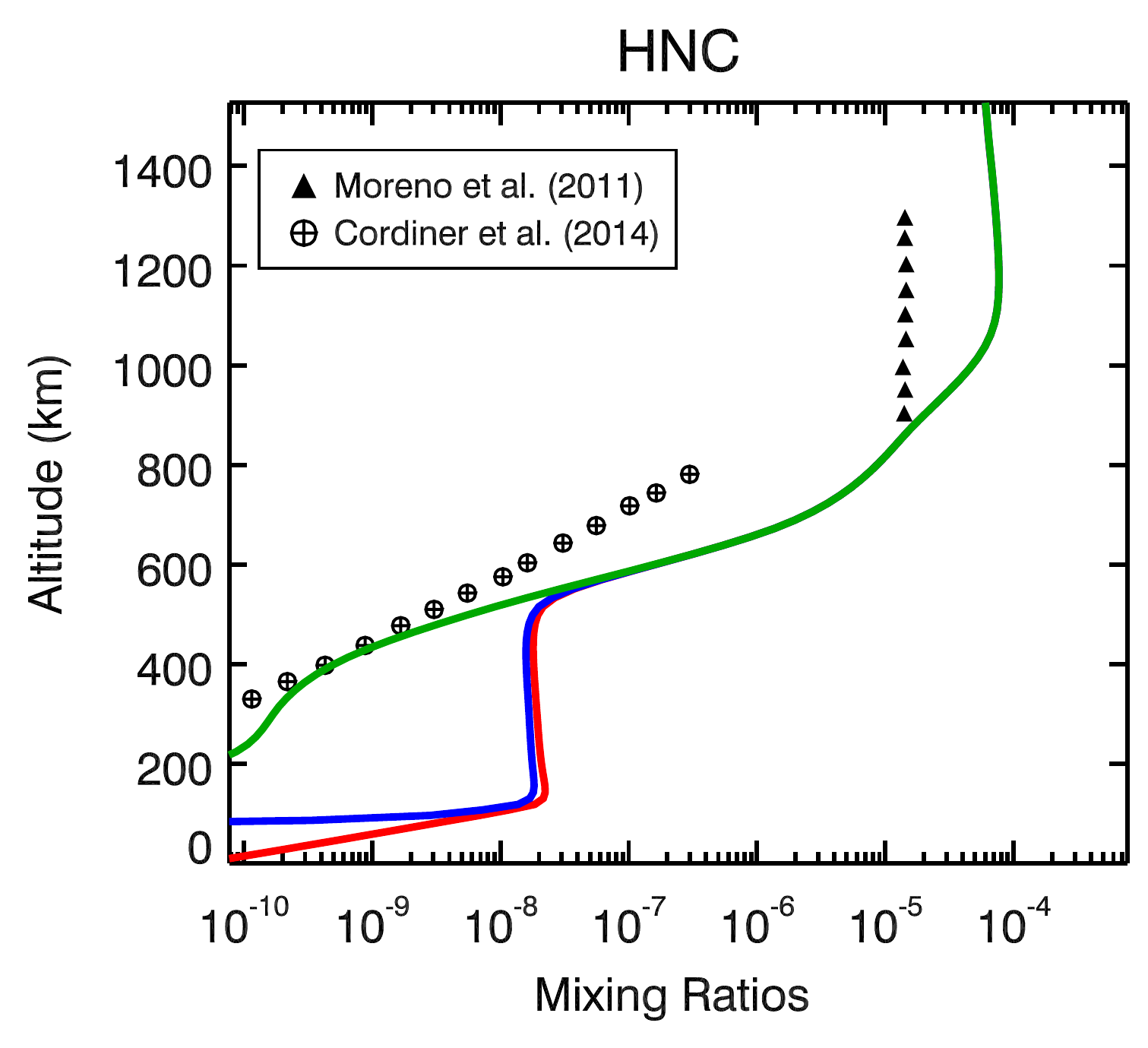}
\includegraphics[width=0.5\linewidth]{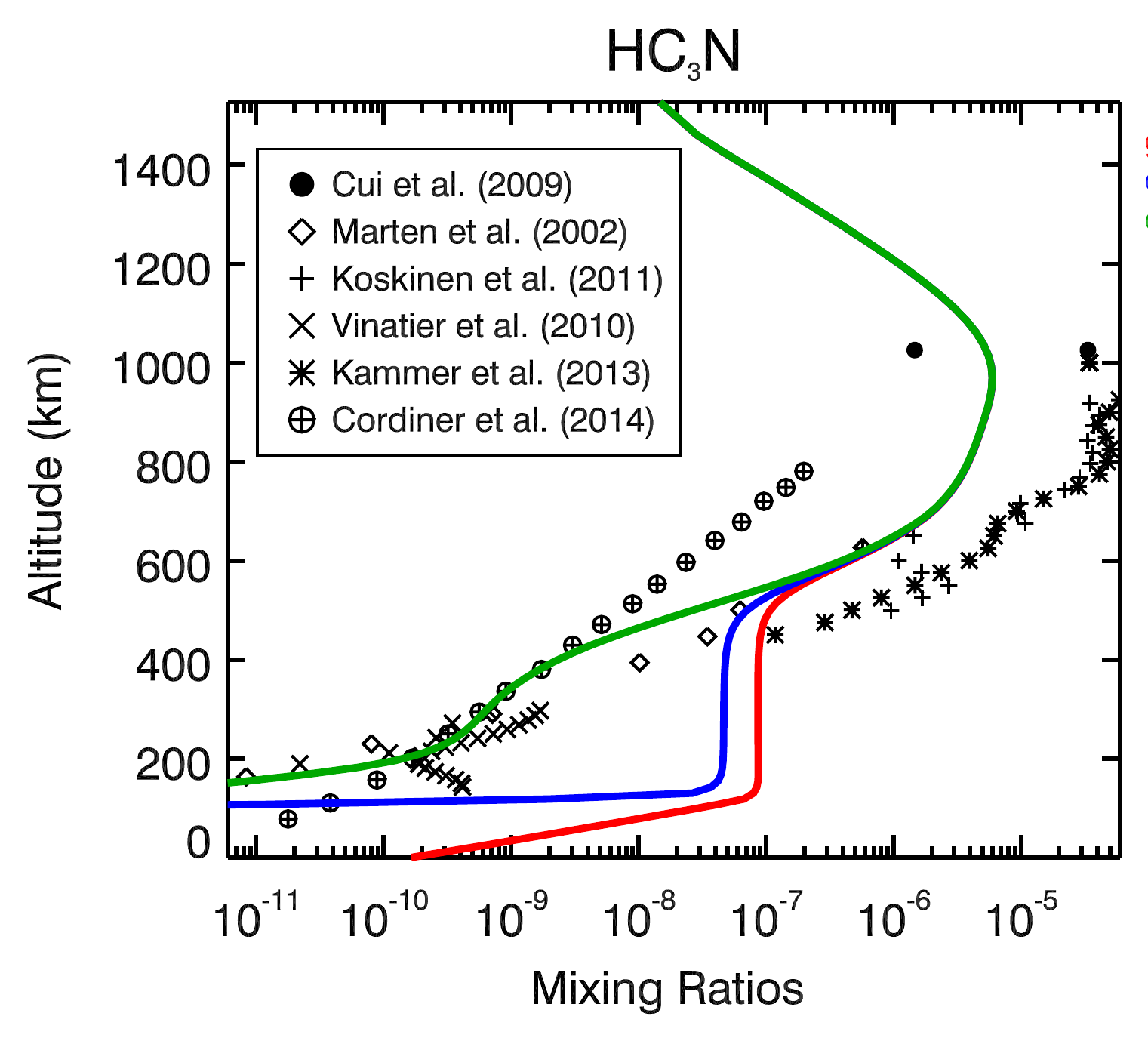}
\caption{\label{fig:nchem1}Distribution of some nitrogen-bearing
  molecules. (red line = Model A, blue line = Model B, green line = Model C).
Cassini/INMS ($\bullet$ \cite{cui09}; $\triangleleft$ \cite{magee09},
$\Leftcircle$ \cite{vuitton07}), Cassini/UVIS (+
\cite{koskinen11}, $\vartriangle$
\cite{shemansky05}, $*$ \cite{kammer15}),
Keck \citep[$\blacktriangledown$][]{kim05}),
Voyager occultation observations ($\square$ \cite{vervack04}),
Cassini/CIRS ($\times$ \cite{vinatier10}, $\ocircle$\cite{nixon10},
$\Diamondblack$ \cite{teanby13}), 
\citep[$\blacktriangle$][]{moreno11}, IRAM
($\diamond$ \cite{marten02}),  ALMA ($\oplus$ \cite{cordiner14}). }
\end{figure}
\begin{figure}
\includegraphics[width=0.5\linewidth]{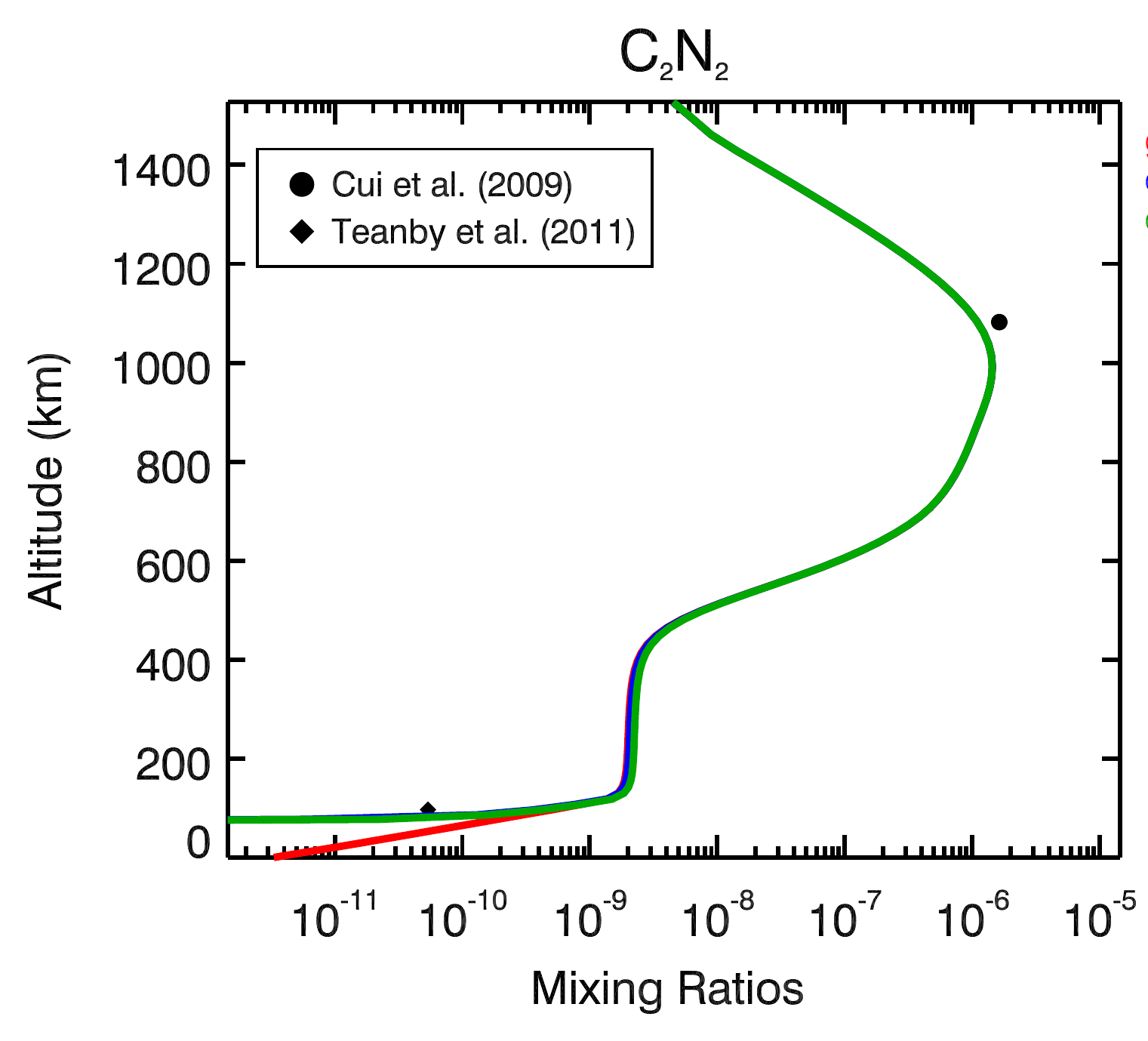}
\includegraphics[width=0.5\linewidth]{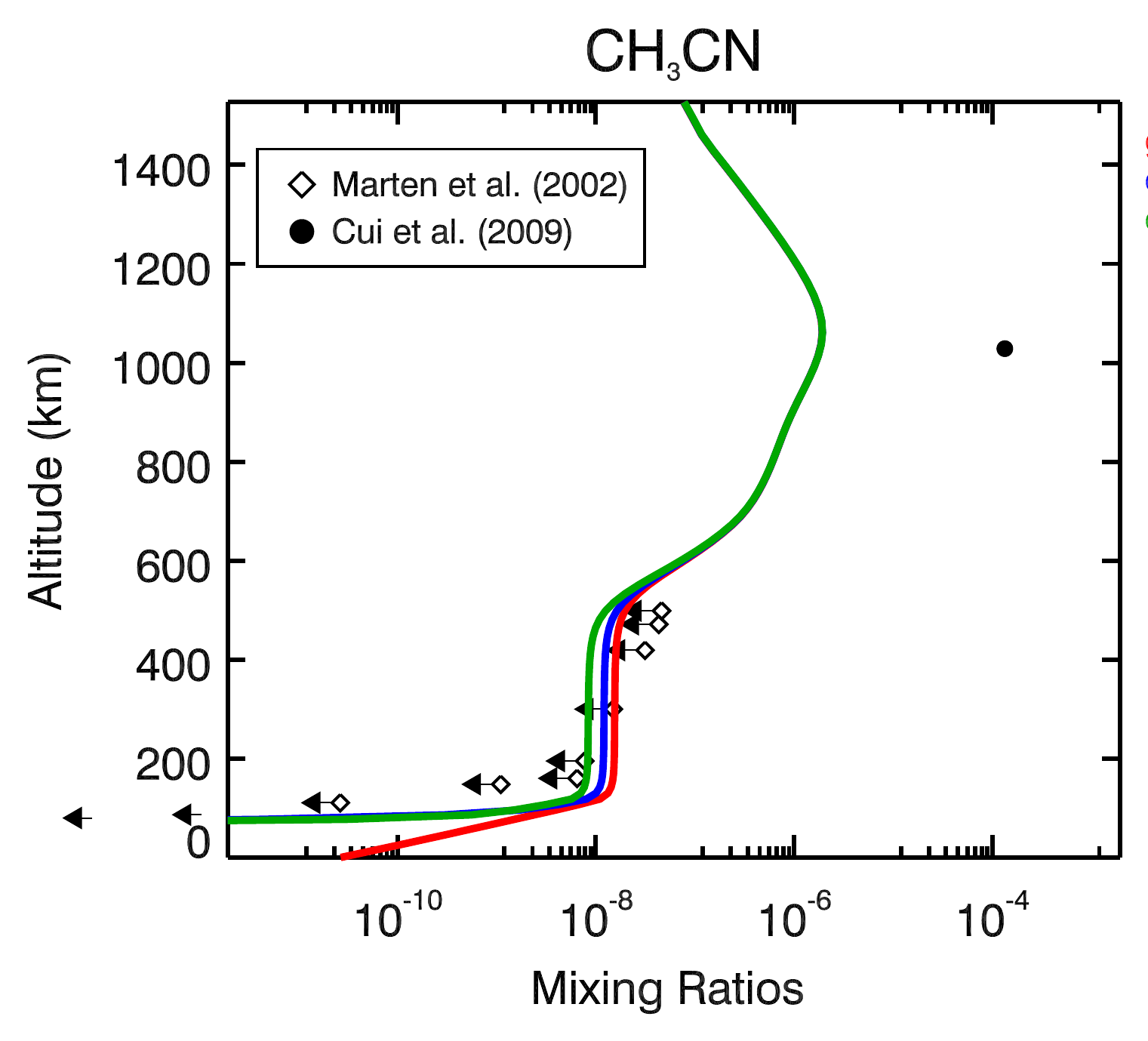}
\includegraphics[width=0.5\linewidth]{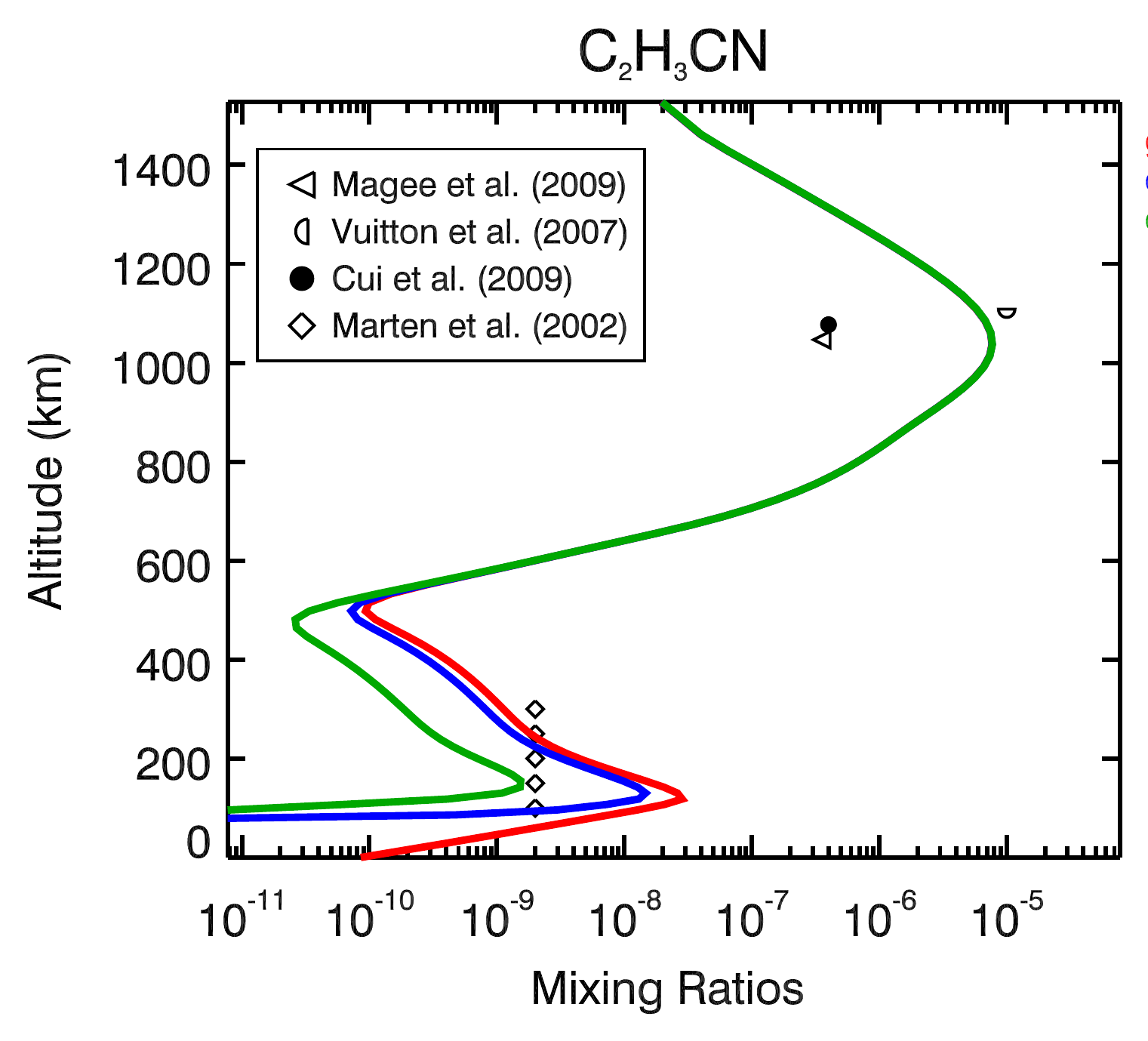}
\includegraphics[width=0.5\linewidth]{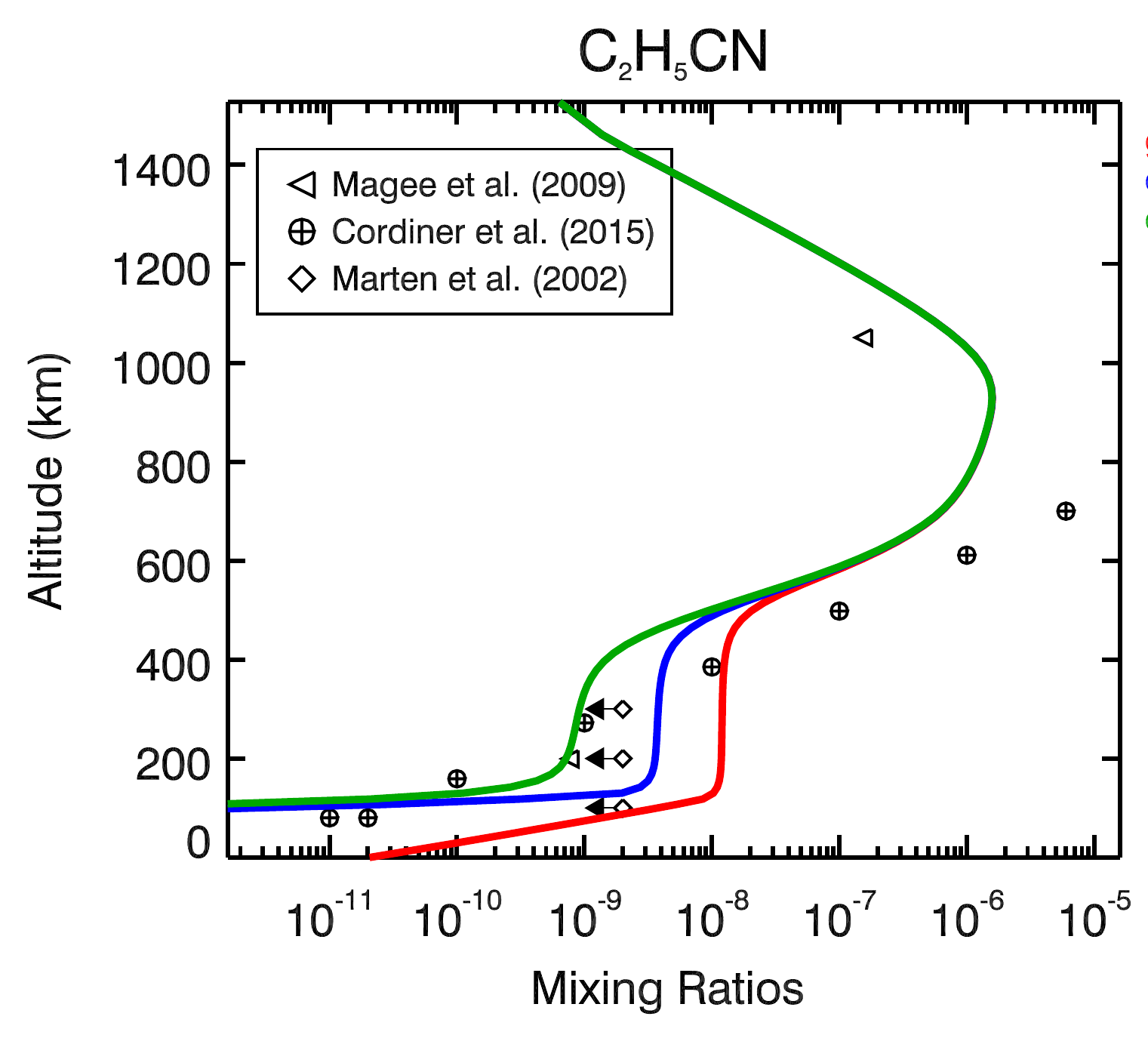}
\caption{\label{fig:nchem2}Abundances of more nitrogen-bearing
  species. Cassini/INMS ($\bullet$ \cite{cui09},
  $\triangleleft$ \cite{magee09}, $\Leftcircle$ \cite{vuitton07}), Cassini/CIRS
  ($\Diamondblack$ \cite{teanby13}),
  IRAM 30m ($\diamond$ \cite{marten02}),
 ALMA data ($\oplus$ \cite{cordiner15}).}
\end{figure}

\subsubsection{NH$_3$}

In the lower atmosphere upper limits of the NH$_3$ abundance are
provided by Herschel/SPIRE measurements \cite[65 -- 100
km;][]{teanby13} and from
CIRS/Cassini limb observations \citep[110 -- 250 km;][]{nixon10}.
In the upper atmosphere the abundance is derived from Cassini/INMS of
NH$_4^+$ \citep{vuitton07} at 1100 km.
\cite{cui09}  claim a detection of NH$_3$ in the ionosphere
between 950 and 1200km.  Their value is an order of magnitude larger than that
derived by Vuitton et al. and its origin is a matter of debate.
It is possible that this high value is due to spent hydrazine fuel \citep{magee09}.

Our model abundances in the upper atmosphere are a factor of 10 lower than 
the observations of \cite{vuitton07} (Figure~\ref{fig:nchem1}).  Below 250 km our models are
considerably lower (but consistent with) the upper limits derived by \cite{teanby13} and \cite{nixon10}.

The main formation processes for NH$_3$ are 
\begin{align}
\hbox{C$_2$H$_3$} + \hbox{NH$_2$} & \longrightarrow \hbox{NH$_3$} +
                                    \hbox{C$_2$H$_2$} & \hbox{$<$ 800
                                                        km} 
\end{align}
with destruction by photodissociation.

As discussed by \cite{loison15} the formation of NH$_3$ via
neutral-neutral reactions depends on the presence of NH$_2$ which is
not efficiently produced in Titan's
atmosphere. The inclusion of ion-molecule chemistry may lead to higher
abundances of NH$_3$.

\subsubsection{HCN}

Observations of HCN have been made from 100 km to 1000 km.  The
millimeter observations of \cite{marten02} covered the whole disk and
were mainly sensitive to the mid-latitude and equatorial regions.
Observations from Cassini/CIRS \citep{vinatier07,vinatier10}, UVIS
\citep{koskinen11,shemansky05,kammer15}, and INMS \citep{magee09}
provide abundance information between 400 and 1000 km.
Abundances in the lower atmosphere are also provided by \cite{kim05}
from Keck observations \citep{geballe03}.  \cite{vervack04} used
Voyager 1 Ultraviolet Spectrometer measurements to determine
abundances between 500 and 900km, although the inferred abundances are
much higher than other estimates.  The differences between the Voyager
1 HCN abundances and those from Cassini may be due to solar cycle
variations.  Investigating such differences is beyond the scope of
this work.

Overall our models are in good agreement with the observational data (Figure~\ref{fig:nchem1}).
We find that condensation and sublimation are important for HCN below 500 km.  The
best fit to the observations is obtained with Model C
(Figure~\ref{fig:nchem1}), where sedimentation and haze formation
reduce the abundance of HCN below 500 km.

The main formation processes are 
\begin{align}
\hbox{HNC} + \hbox{H} & \longrightarrow \hbox{HCN} + \hbox{H} & \hbox{300 -- 800 km}\\
\hbox{CN} + \hbox{CH$_4$} & \longrightarrow \hbox{HCN} + \hbox{CH$_3$} & \hbox{200 -- 600 km}\\
\hbox{N} + \hbox{CH$_2$} & \longrightarrow  \hbox{HCN} + \hbox{H} & \hbox{600 -- 900 km}\\
\hbox{C$_2$H$_3$CN} + h\nu & \longrightarrow \hbox{HCN} + \hbox{C$_2$H$_2$} & \hbox{$<$ 1000 km}\\
\hbox{H$_2$CN} + \hbox{H} & \longrightarrow \hbox{HCN} + \hbox{H}_2 & \hbox{900 - 1300 km}
\end{align}

Photodissociation plays a role in both the formation of HCN (via
photodisssociation of C$_2$H$_3$CN above 1000 km) and in its
destruction (forming CN and H).  Below 200 km destruction is by
\begin{equation}
\hbox{C$_2$H$_3$} + \hbox{HCN} \longrightarrow \hbox{C$_2$H$_3$CN} + \hbox{H}
\end{equation}

\subsubsection{HNC}

The first observations of HNC in Titan were made using Herschel/HIFI
by \cite{moreno11}.  These measurements do not allow the exact
vertical abundance profile to be determined.  Several
possible profiles can fit the data depending on the mixing ratio and
the cut-off altitude assumed.  \cite{loison15} suggest two possible
profiles: one where the mixing ratio of HNC is 1.4 \x 10$^{-5}$ above 900
km (shown in Figure~\ref{fig:nchem1}) and another where the mixing
ratio is 6 \x 10$^{-5}$ above 1000 km.  Our models fall between these two ranges.

More recently \cite{cordiner14} used ALMA to detect HNC.  They found
that the emission mainly originates at altitudes above 400 km and that
there are two emission peaks that are not symmetrical in longitude.
We are able to match their best fit profile reasonably well with model C 
(green line; Figure~\ref{fig:nchem1}),  where HNC
forms haze providing the best agreement with the data at lower altitudes. 

The main formation channels of HNC are
\begin{align}
\hbox{C$_2$H$_3$CN} + h\nu & \longrightarrow \hbox{HNC} + \hbox{C$_2$H$_2$} &
                                                                \hbox{$<$
                                                                900 km}\\
\hbox{H$_2$CN} + \hbox{H} & \longrightarrow  \hbox{HNC} + \hbox{H$_2$} & \hbox{$>$
                                                           900 km}\\
\hbox{N} + \hbox{CH$_3$} & \longrightarrow \hbox{HNC} + \hbox{H} & \hbox{$>$ 900 km}
\end{align}

\begin{figure}
\centering
\includegraphics[width=0.5\linewidth]{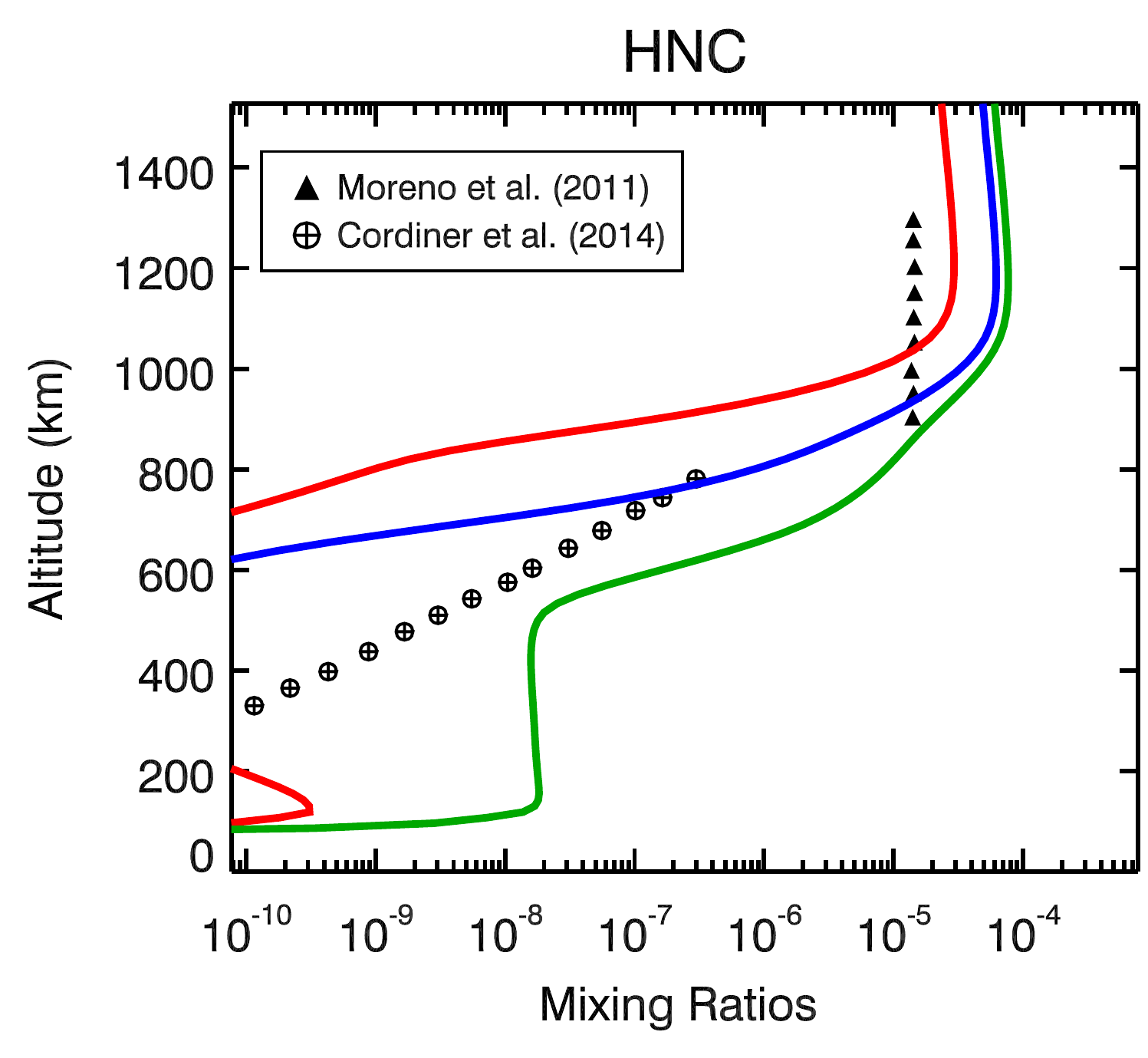}
\caption{\label{fig:HNCcomp}How the HNC abundance depends on the 
activation barrier of the reaction H + HNC $\longrightarrow$ HCN + H. 
{\it green} E$_A$ = 2000 K \citep{kida}, {\it blue} E$_A$ = 1400 K, {\it red}
E$_A$ = 800 K \citep{loison15}. The lower
activation barrier results in more HNC being converted into HCN 
but does not result in a better agreement with the altitude distribution seen in the 
ALMA observations.}
\end{figure}

The main destruction process is by reaction with H atoms forming HCN.
This reaction has an activation barrier.  In the literature the value for the activation barrier
ranges from 800 to 2000 K \citep{te96,sn98,petrie02,kida}.  Here we are using the rate
from the KIDA database \citep{kida} which has the highest activation barrier of 2000 K.  
\cite{loison15} used the lowest value (800 K), resulting in more efficient HCN production 
and consequently a lower gas phase abundance of HNC
than we see here.  We find that reducing the activation barrier does indeed reduce the
mixing ratio of HNC but does not result in a good fit to the ALMA observations in this region 
(Figure~\ref{fig:HNCcomp}).

\subsubsection{HC$_3$N}
HC$_3$N has been observed at altitudes from 200 to 1000 km
\citep{marten02,vervack04,teanby06,vuitton07,cui09,magee09,vinatier10,cordiner14}.
Below 500 km our models are in excellent agreement with the
observations if it is assumed that HC$_3$N forms aerosols and thus is
permanently removed from the gas 
(Figure~\ref{fig:nchem1}{\it ~(bottom left)}).  Condensation and sublimation alone
result in an over-estimate of the abundance compared to the
observations in this region.  
Good agreement is also seen for all models between 500 km
and 700 km.  Above this our models tend to under predict the HC$_3$N
abundance.
 Below 100 km the mixing ratio follows the
saturation level, so that below this
altitude the mixing ratio is much reduced compared to the gas only model.   Better
agreement with the observations below 400 km is obtained in the
haze formation model where condensed molecules are assumed to be
incorporated into aerosol particles and removed from the gas.

The main formation process below 1000 km is
\begin{equation}
\hbox{C$_3$N} + \hbox{CH$_4$} \longrightarrow \hbox{HC$_3$N} + \hbox{CH}_3
\end{equation}
and above 800 km by
\begin{align}
\hbox{CN} + \hbox{C$_2$H$_2$} &  \longrightarrow \hbox{HC$_3$N} + \hbox{H} \\
\hbox{C$_2$H$_3$CN} + h\nu & \longrightarrow \hbox{HC$_3$N} + \hbox{H}_2
\end{align}
Destruction is by photodissociation
\begin{equation}
\hbox{HC$_3$N} + h\nu \longrightarrow \hbox{C$_3$N} + \hbox{H}
\end{equation}
and by reaction with H atoms
\begin{equation}
\hbox{H} + \hbox{HC$_3$N} + \hbox{M} \longrightarrow \hbox{H$_2$C$_3$N} + \hbox{M}
\end{equation}

The observations show a sharp decrease in the abundance of HC$_3$N below
400 km.  In our models this can be accounted for if HC$_3$N is incorporated
into haze particles (Model C).  An alternative explanation of
meridional circulation and condensation in the polar regions has been
suggested \citep{loison15,hourdin04}.

\subsubsection{C$_2$N$_2$}
Observations of C$_2$N$_2$ have been made by Cassini/CIRS \citep{teanby06,teanby09} and by
Cassini/INMS \citep{cui09,magee09}.  The models with condensation are
in very good agreement with both of these datasets (Figure~\ref{fig:nchem2}).  Without
condensation the abundance in the lower atmosphere is over-estimated.  

The main formation route for C$_2$N$_2$ is by the reaction of CN and HNC:
\begin{equation}
\hbox{CN} + \hbox{HNC} \longrightarrow \hbox{C$_2$N$_2$} + \hbox{H}
\end{equation}
with destruction via photodissociation forming CN or by 
\begin{equation}
\hbox{H} + \hbox{C$_2$N$_2$} + \hbox{M} \longrightarrow \hbox{H$_2$CN$_2$} + \hbox{M} \hbox{~~~~$<$ 400 km}
\end{equation}

\subsubsection{CH$_3$CN}

Submillimeter observations with the IRAM 30 m telescope detected
the CH$_3$CN (12-11) rotational line providing a disk average vertical
profile up to 500 km, dominated by the equatorial region \citep{marten02}.  
Cassini/CIRS \citep{nixon13} and Cassini/INMS
\citep{vuitton07,cui09} provide estimates of the abundance above 1000 km.

All models are in good agreement with the observations below 800 km, although all predict
slightly lower abundances than observed between 500 and 600 km.  The predicted mixing ratio
at 1100 km is a factor of 10 lower than the observed value of 3 \x 10$^{-5}$ \cite{cui09}.

The main formation processes are
\begin{align}
\hbox{CN} + \hbox{CH}_4 & \longrightarrow  \hbox{CH$_3$CN} + \hbox{H} & \hbox{$<$ 900 km}\\
\hbox{NH} + \hbox{C$_2$H$_4$} & \longrightarrow  \hbox{CH$_3$N} + \hbox{H}
                                  &  \hbox{$>$ 900 km}\\
\hbox{CH$_2$NH} + \hbox{CH} & = \hbox{CH$_3$CN} + \hbox{H}  & \hbox{$<$ 900 km}
\end{align}
with destruction by
\begin{align}
\hbox{CH$_3$CN} + \hbox{H} & \longrightarrow  \hbox{CN} + \hbox{CH}_4 & \hbox{$<$ 1200 km} 
\end{align}

\subsubsection{C$_2$H$_3$CN}

Several observations have placed upper limits on the abundance of
C$_2$H$_3$CN. \cite{marten02} used the IRAM 30m telescope to determine
upper limits between 100 and 300 km. Cassini/INMS has
provided an upper limit of 4 \x 10$^{-7}$ at 1077 km \citep{cui09},
while Cassini/INMS \cite{magee09} determined a mixing ratio of 3.5 \x
10$^{-7}$ at 1050 km and \cite{vuitton07} found 10$^{-5}$ at 1100 km
from observations of ions.  Cordiner et al. (private communication) 
have detected C$_2$H$_3$CN in the submillimeter and found an average
abundance of 1.9 x 10$^{-9}$ above 300 km.The model abundance of C$_2$H$_3$CN in the
upper atmosphere is within a factor of 2 of the \cite{vuitton07} value
but 50 times higher than \cite{magee09} and \cite{cui09}.  

None of our models have a constant mixing ratio with altitude between
100 and 300 km as derived from the IRAM observations (Figure~\ref{fig:nchem2}).  Model B and C
(which include condensation) are consistent with the derived mixing
ratio at a particular altitude, but neither reproduce the constant
value between 100 and 300 km.  In the upper atmosphere all models
predict mixing ratios within a factor a 3 of the \cite{magee09} result
but are over-abundant compared to the other measurement in this region.

The main production mechanism is by
reaction of CN with C$_2$H$_4$:
\begin{align}
\hbox{C$_2$H$_3$} + \hbox{HCN} & \longrightarrow \hbox{C$_2$H$_5$CN} + \hbox{H} & \hbox{$<$ 200 km}\\
\hbox{CN} + \hbox{C$_2$H$_4$} & \longrightarrow \hbox{C$_2$H$_3$CN} + \hbox{H} &
                                                                   \hbox{400 -- 800
                                                                   km}\\
\hbox{H$_2$C$_3$N} + \hbox{H} & \longrightarrow \hbox{C$_2$H$_3$CN} & \hbox{$>$ 400 km}
\end{align}

Gas phase destruction processes are
\begin{align}
\hbox{C$_2$H$_3$CN} + h\nu & \longrightarrow \left\{
\begin{array}{ll} \hbox{HCN} + \hbox{C$_2$H$_2$}\\
\hbox{HNC} + \hbox{C$_2$H$_2$}\\
\hbox{HC$_3$N} + \hbox{H}_2
\end{array}
\right.
\end{align}

Below 400 km haze formation and sedimentation of aerosol particles
play an important role in determining the gas mixing ratio in Model C.

\subsubsection{C$_2$H$_5$CN}

Upper limits for the abundance of C$_2$H$_5$CN have been determined between 100 and 300
km from IRAM 30m observations \cite{marten02}, with abundances in the upper
atmosphere provided by Cassini/INMS data \citep{vuitton07,cui09,magee09}.
More recently
\cite{cordiner15} detected this molecule above 200 km using ALMA. 

Our models over-estimate the abundance of C$_2$H$_5$CN in the upper
atmosphere (Figure~\ref{fig:nchem2}), probably because we do not include ion chemistry \citep[for
a discussion of this point see][]{loison15}.  Below 700 km, Model C
is in  excellent agreement with the ALMA data of Cordiner et al.

The main formation process is
\begin{equation}
\hbox{CH$_2$CN} + \hbox{CH$_3$} + \hbox{M} \longrightarrow \hbox{C$_2$H$_5$CN} + \hbox{M}
\end{equation}
Destruction is by photodissociation
\begin{align}
\hbox{C$_2$H$_5$CN} + h\nu & \longrightarrow \left\{
\begin{array}{ll}
\hbox{C$_2$H$_4$} + \hbox{HCN} \\
\hbox{CH}_3 + \hbox{CH$_2$CN} \\
\hbox{C$_2$H$_3$CN} + \hbox{H$_2$}
\end{array}
\right.
\end{align}
or by reaction with CH, C$_2$H$_3$ or C$_2$H.
\begin{align}
\hbox{C$_2$H$_5$CN} + \hbox{CH} & \longrightarrow \hbox{CH$_2$CN} + \hbox{C$_2$H$_4$}\\
\hbox{C$_2$H$_5$CN} + \hbox{C$_2$H} & \longrightarrow \hbox{HC$_3$N}  + \hbox{C$_2$H$_5$} & \hbox{$<$ 300 km}\\
\hbox{C$_2$H$_3$CN} + \hbox{C$_2$H$_3$} & \longrightarrow \hbox{C$_2$H$_3$CN} + \hbox{C$_2$H$_5$} & \hbox{$<$ 300 km}
\end{align}

\subsection{\label{sec:flux}Condensates in Titan's Atmosphere}

\begin{figure}
\includegraphics[width=0.5\linewidth]{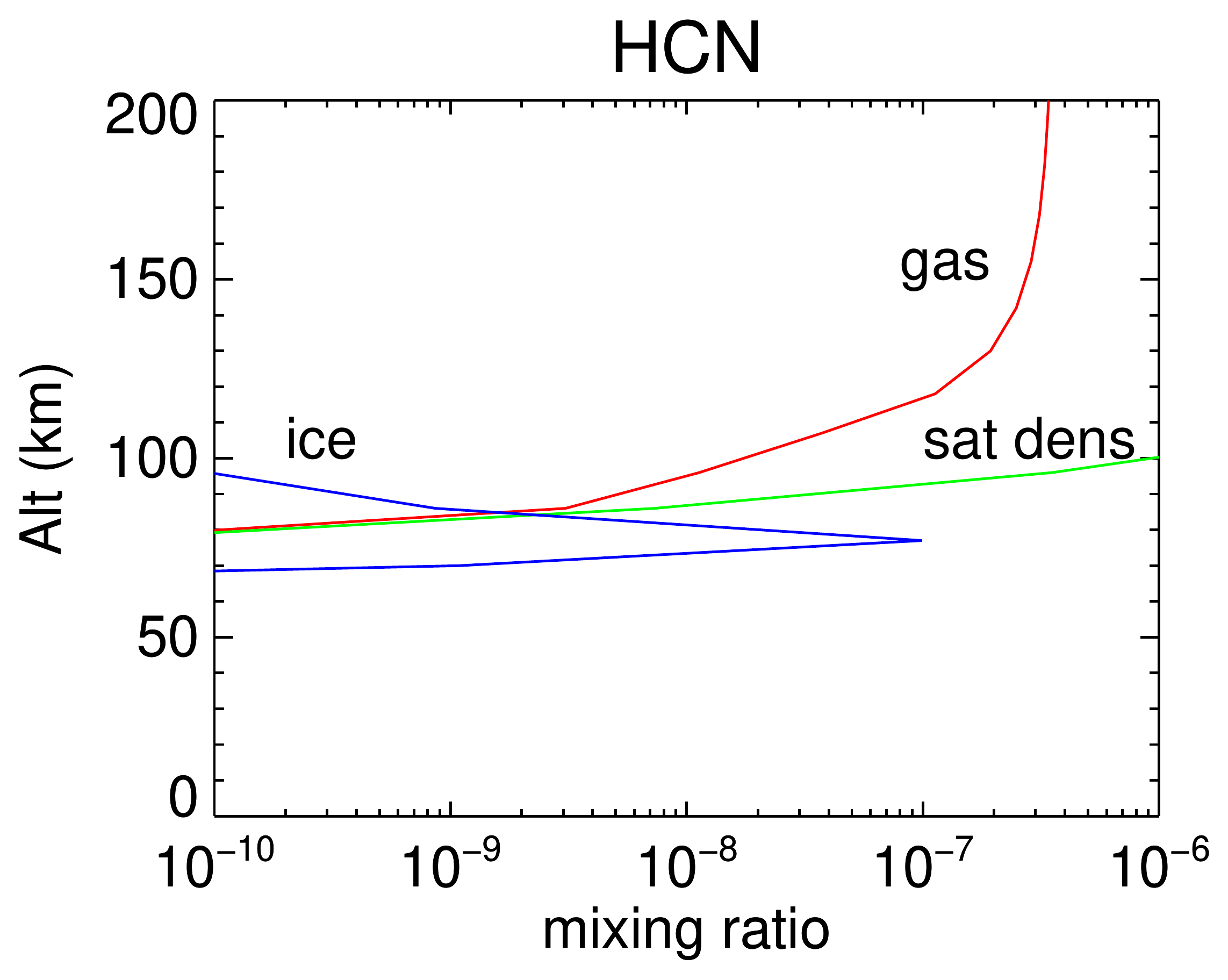}
\includegraphics[width=0.5\linewidth]{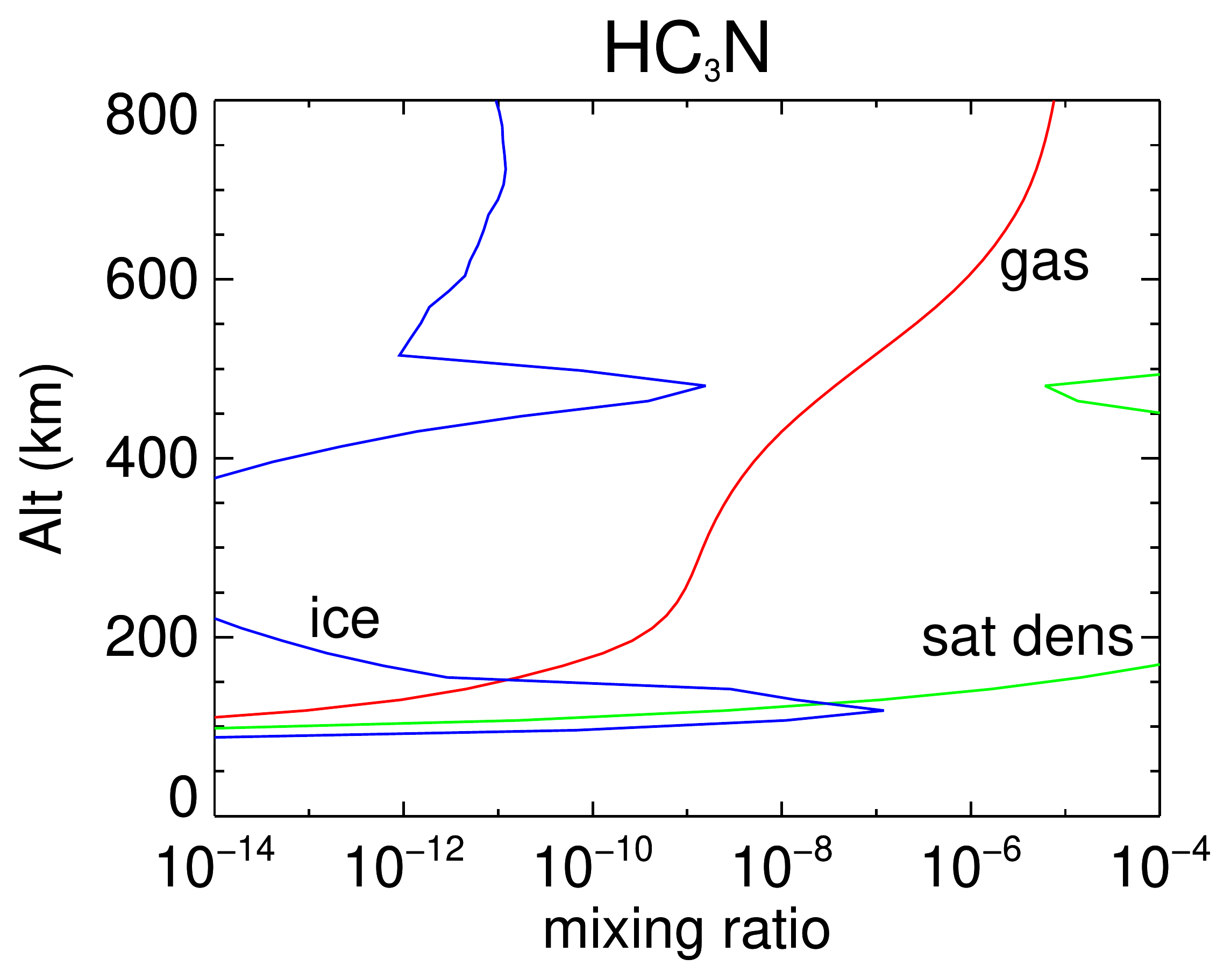}
\caption{\label{fig:HCN_vap}The calculated mixing ratios of HCN (left)
  and HC$_2$N (right) in the
  gas (red line) and in condensed form (blue line).  Also shown is the
  saturated mixing ratio (green line).  The mixing ratio of HCN$^c$
  peak at $\sim$ 65 km.  The mixing ratio of HCN in the gas around this
  altitude is  slightly below the saturated value, even though there
  is condensed HCN available to be sublimated.  The reason for this is
  that the condensates are not pure.  Since the sublimation rate
  depends on the surface coverage of the condensed molecule
  sublimation from a mixed condensate is less efficient than from a
  pure condensate, leading to lower gas abundances.  A similar effect
  is seen for other condensed molecules.  For HC$_3$N the peak
  condensate mixing ratio occurs around 500 km where there is a dip in
temperature corresponding to a high local abundance of HC$_3$N.  Other
nitrogen-bearing molecules show similar behavior in this region.}
\end{figure}

We find several layers at which condensates are abundant with the location
being molecule dependent.  The first condensate layer
is in the lower atmosphere around the tropopause. Here we find
condensates of C$_2$H$_2$, C$_2$H$_4$, C$_2$H$_6$, C$_3$H$_8$ among others.
 A little further up in the atmosphere around 65 -- 80 km several molecules have peaks in
condensation e.g.\ HCN, C$_4$H$_8$, C$_4$H$_2$, C$_2$H$_3$CN, C$_2$N$_2$,
CH$_3$C$_2$H.  Another layer of C$_2$H$_3$CN, CH$_3$CN, C$_2$H$_5$CN
and CH$_3$C$_2$H forms around 110 -- 130 km.  Several molecules also
have high condensation levels between 600 and 900 km
e.g. CH$_3$C$_2$H, HC$_5$N, HC$_3$N, CH$_3$CN, and
C$_2$H$_5$CN. Figure~\ref{fig:HCN_vap} shows the condensation layers for
HCN and HC$_3$N.  Both these molecules have high condensate abundances
between 70 and 100 km, but HC$_3$N has a further peak around 500 km where
the atmospheric temperature dips, and the gas phase abundance of this molecule is high.

The net flux of material falling on to the surface of Titan can be
calculated from the difference between the atmospheric formation and
destruction.  
Table~\ref{tab:flux} presents our predictions of the surface flux of
nitrogen molecules.  These are in solid form and if evenly distributed
across Titan's surface would create a layer 4.4 m deep over a timescale
of 1 Gyr.  This amount of ``fixed nitrogen'' could be of biological importance.

\begin{table}
\small
\centering
\begin{tabular}{lcrrc}
\hline
Molecule & Flux & Flux & Depth (m) & Solid/Liquid  \\
     & (molecules cm$^{-2}$ s$^{-1}$) &(g cm$^{-2}$/Gyr) & &  at 95 K \\
\hline
HCN$^c$  & -1.2 \x 10$^8$ & -170 & 2.12 & S  \\
HNC$^c$ & -8.1 \x 10$^6$ & -11.0 & 0.14 & S\\
HC$_3$N$^c$  & -2.9 \x 10$^7$ & -77.4  & 1.0 & S \\
HC$_5$N$^c$ & -3.4 \ 10$^6$ & -13.5 & 0.17\\
C$_2$N$_2^c$  & -5.8 \x 10$^5$ & -15.8 & 0.02 & S \\
CH$_3$CN$^c$  & -2.8 \x 10$^5$ & -0.6 & 0.01 & S \\
CH$_3$C$_2$CN & -4.5 \x 10$^6$ & -15.3 & 0.19 & \\
C$_2$H$_5$CN$^c$  &  -6.4 \x 10$^6$ & -18.4 & 0.23  & S \\
C$_2$H$_3$CN$^c$  &  -1.5 \x 10$^7$ & -41.6 & 0.52 & S  \\
\hline
Total N & & & 4.4 m & \\
\hline
\end{tabular}
\caption{\label{tab:flux}Flux of condensed molecules onto
  Titan's surface.  Also shown are the estimated deposit thickness
  (depths) calculated
assuming an average density of 0.8 g cm$^{-3}$.}
\end{table}

\section{Discussion and Conclusions}\label{sec:conc}

The removal of molecules by condensation plays an important role in determining 
the gas phase composition of Titan's atmosphere, as well as creating 
new aerosols.  Condensates are found throughout the atmosphere. 
For the majority of molecules, condensation is most efficient below the tropopause.
Larger molecules, and in particular nitrogen-bearing molecules
have another condensation peak between 200 and 600 km.
Relatively high abundances of condensates can also be present above
500 km if the gas phase abundance of a given molecule is high, e.g.\
HC$_3$N, HC$_5$N, CH$_3$CN and C$_2$H$_5$CN.  These molecules
condense in the region where Titan's haze forms. The
effect is enhanced if it is assumed that some molecules can be
permanently removed from the gas by being incorporated into aerosol
particles.  This mechanism was able to bring the abundances of
HC$_3$N, HCN, HNC, CH$_3$CN and C$_2$H$_5$CN into good 
agreement with the observations below 600 km.

Although Titan possesses a rich organic chemistry it is unclear whether this could
lead to life.  Photochemically produced compounds on Titan, 
principally acetylene, ethane and organic solids, would release energy 
when consumed with atmospheric hydrogen, which is also a photochemical 
product. \cite{mckaysmith05} speculate on the possibility of widespread 
methanogenic life in liquid methane on Titan. On Earth fixed nitrogen is often 
a limiting nutrient. Our work shows that an abundant supply of fixed nitrogen, 
including species of considerable complexity, is available from atmospheric photochemistry.

Creating the kinds of lipid membranes that form the basis of lie on Earth
depends on the  presence of liquid water.  Titan's atmosphere contains little oxygen
and the surface temperature is well below that at which liquid water can survive. 
Instead surface liquids are hydrocarbons \citep[e.g.][]{hayes16}.  Therefore any
astrobiological processes, if present, are likely to be quite different to those on Earth.
A recent paper by \cite{slc15} suggests that as alternative to lipids, membranes could
be formed from small nitrogen-bearing organic molecules such as acrylonitrile (C$_2$H$_3$CN).
Stevenson et al.\ calculate that a membrane composed of acrylonitrile molecules would
be thermodynamically stable at cryogenic temperatures and would have a high energy barrier
to decomposition.

All of our models predict abundances of C$_2$H$_3$CN that are in agreement with observations
above 500 km.  Below this condensation and incorporation into haze are required to bring the
predicted mixing ratios down to the values inferred from observations \cite{cordiner15}.  If
acrylonitrile were to be involved in life formation it needs to reach the surface of Titan.  
Our predicted flux of this molecule onto Titan's surface is 1.5 \x 10$^7$ molecules cm$^{-2}$ s$^{-1}$, 
or $\sim$ 41.5 gcm$^{-2}$/Gyr, a quantity that is potentially of biological importance.

\section*{Acknowledgements}
This research was conducted at the Jet Propulsion
Laboratory, California Institute of Technology under contract with the
National Aeronautics and Space Administration.  
Support was provided by the NASA Astrobiology Institute/Titan as a Prebiotic
Chemical System.  YLY was supported in part by the Cassini UVIS program via NASA grant
JPL.1459109 to the California Institute of Technology. The
authors thank Dr. Run-Lie Shia for his assistance with the KINETICS code
and Dr. Panyotis Lavvas for providing the aerosol data used in these models.

%\software{KINETICS; \citep{allen81}}

\bibliography{titan}
\end{document}